\documentclass{aa}  
\usepackage{graphicx}
\usepackage{txfonts}
\usepackage[dvipsnames]{xcolor}
\usepackage{amssymb}
\usepackage{siunitx}
\usepackage{soul}
\begin{document}

   \title{Complete X-ray census of M\,dwarfs in the solar Neighborhood}

      \subtitle{I. GJ\,745\,AB: Coronal-hole Stars in the $10$\,pc Sample}
\author{M. Caramazza
          \inst{1}
          \and
          B. Stelzer\inst{1,2}
          \and
          E. Magaudda\inst{1}
          \and
          St. Raetz\inst{1}
          \and
          M. G\"udel \inst{3} \and S. Orlando \inst{2}
          \and K. Poppenh\"ager\inst{4,5} 
          }

   \institute{$^1$Institut für Astronomie und Astrophysik, Eberhard-Karls Universität Tübingen, Sand 1, 72076 Tübingen, Germany\\
   $^2$INAF–Osservatorio Astronomico di Palermo, Piazza del Parlamento 1,90134 Palermo, Italy\\
   $^3$University of Vienna, Department of Astrophysics, Türkenschanzstrasse 17, 1180 Vienna, Austria\\
   $^4$Leibniz Institute for Astrophysics Potsdam (AIP), An der Sternwarte 16, D-14482 Potsdam, Germany\\ 
   $^5$Universität Potsdam, Institut für Physik und Astronomie, Karl-Liebknecht-Straße 24/25, D-14476 Potsdam, Germany \newline
    \email{caramazza@astro.uni-tuebingen.de}}

   \date{Received 21-03-2023; accepted 22-05-2023}

 
  \abstract
   { X-ray emission is the most sensitive  diagnostic of magnetic activity in M\,dwarfs and, hence, of the dynamo in low-mass stars. Moreover it is crucial to 
   quantify the
   influence of the
   stellar irradiation on the evolution of planet atmospheres.}
   {We have embarked in a systematic study of 
   the X-ray emission in a volume-limited sample of M dwarf stars, in order to explore the full range of activity levels present in their coronae and, thus, to obtain a better understanding of the conditions in their outer atmospheres
   and their possible impact on the circumstellar environment.}
   {We identify in a recent catalog of the {\it Gaia} objects within $10$\,pc from the Sun 
   all stars with spectral type between M0 and M4, and search   systematically 
   for X-ray measurements of this sample. To this end, we use both
   archival data (from ROSAT, {\it XMM-Newton}, and from the ROentgen Survey with an Imaging Telescope Array (eROSITA) onboard the Russian Spektrum-Roentgen-Gamma mission) and our own dedicated {\it XMM-Newton}  observations. To make inferences on the properties of the M dwarf corona we compare the range of their observed X-ray emission levels to the flux radiated by the Sun from different types of magnetic structures: coronal holes, background corona, active regions and cores of active regions. In this work we focus on the properties of the stars with the  faintest X-ray emission.    }
   {At the current state of our project, with more than $90$\,\% of the $10$\,pc M dwarf sample observed in X-rays, only one star, GJ\,745\,A, has no detection. 
   With an upper limit luminosity of $\log{L_{\rm x}}\,{\rm[erg\ s^{-1}] } < 25.4$ and an X-ray surface flux of $\log{F_{\rm X,SURF}\,[{\rm erg\ cm^{-2}\ s^{-1}}] < 3.6}\, $
   GJ\,745\,A defines
   the lower boundary of the X-ray emission level of M\,dwarfs. Together with its proper motion companion GJ\,745\,B, GJ\,745\,A it is the only star in this volume-complete sample located 
   in the range of X-ray surface flux 
  that corresponds to the faintest solar coronal structures, the coronal holes. The ultra-low X-ray emission level of GJ\,745\,B ($\log{L_{\rm x}\,{\rm [erg\ s^{-1}]}}={\rm25.6}$ and $\log{F_{\rm X,SURF}\,{\rm [erg\ cm^{-2}\ s^{-1}]} = 3.8}\, $) 
  is entirely attributed to flaring activity, indicating that, while its corona is dominated by `holes',  
  at least one magnetically active structure is present that determines the 
  total X-ray brightness  and the coronal temperature of the star.}
   {}


   \maketitle
%

\section{Introduction}\label{sect:introduction} 

  M\,dwarfs are the most abundant stars in the Galaxy \citep{Chabrier01.0}. They also constitute the majority of the hosts of small, rocky planets \citep{Howard12.0} with estimates for the  occurrence rate for Earth-like planets among early-M\,dwarfs ranging from  0.10 to 0.85 \citep[e.g.][]{Dressing13.0, Pinamonti22.0} and with up to $\approx 25$\,\% of these planets being considered habitable \citep{Pinamonti22.0}. The characterization of the M dwarf population is, therefore, of importance for our understanding of both stellar evolution and the variety of exoplanet systems.

One piece in this puzzle is the high-energy emission from the stellar corona. In analogy with our Sun, the outer atmosphere of M\,dwarfs is considered to be heated by magnetic processes to temperatures above a million Kelvin that provide thermal emission in the UV and X-ray regime. X-ray emission is the most sensitive tracer for magnetic activity in M dwarf stars \citep{Stelzer13.0}. 
The stellar dynamo that underlies these high-energy phenomena is driven by convection and (differential) rotation \citep{Parker_55,Parker_75}. Both these parameters change across stellar mass and evolution, leading to a broad range of activity levels at given mass or age. Despite numerous studies of the subject \citep{Pallavicini_81, Barbera_93, Fleming_95, Schmitt_97, Fleming_98, Marino_00}, the full range of activity levels exhibited by low-mass stars has not yet been fully explored. The key to solving this problem are volume-limited samples.

 X-ray luminosity functions for  volume-limited samples of field dwarf stars have before been presented by \cite{Schmitt95.0} and \cite{Schmitt04.0} based on ROSAT observations. The latter study comprised $37$ stars with spectral types (SpT) M0...M4 within $6$\,pc. Since then, astrometric  surveys have provided updates to the census of X-ray properties of the solar neighborhood.  \cite{Stelzer13.0} have used the {\sc superblink} proper motion survey by \cite{Lepine11.0} to study the X-ray emission of nearby M\,dwarfs.
Complementing ROSAT all-sky survey data with archival (serendipitous) {\it XMM-Newton} observations it was found that $\sim 40$\,\% of the M\,dwarfs within $10$\,pc of the Sun still had no sensitive limit on their X-ray luminosities. This has led us to embark into a dedicated {\it XMM-Newton} program to complete the $10$\,pc M dwarf X-ray luminosity function. In the meantime, improved astrometry has been provided through {\it Gaia} and a new census of nearby stars was published, the {\it `$10$\,parsec  sample in the Gaia era'} \citep{Reyle_2021}. We use this updated $10$\,pc sample as a basis of our effort to provide an unbiased characterization of M\,dwarfs in the X-ray band.

 With a systematic compilation from X-ray archives integrated by our dedicated observations to complete the M dwarf $10$\,pc X-ray census we are, for the first time, able to probe the full range of X-ray luminosities ($L_{\rm x}$) present in early-M dwarf stars (spectral type M0$-$M4). We find that their X-ray activity levels span three orders of magnitude, from the canonical saturation level of $\log{(L_{\rm x}/L_{\rm bol} \approx -3)}$ or surface flux $\log{F_{\rm X,SURF}}\,{\rm [erg/cm^2/s]} \approx 7$, 
 and higher during flares, to $\log{F_{\rm X,SURF}}\,{\rm [erg/cm^2/s]} \approx 4$.
 Remarkably, there is one star in the $10$\,pc sample, the binary GJ\,745\,AB, that appears to have an X-ray emission level significantly below this lower bound. Assuming that M dwarf coronae are composed of the same types of magnetic structures as the Sun, the only feature that can explain such low-level X-ray emission are coronal holes (CH). Coronal holes are regions of the solar corona characterized by low density plasma associated to open magnetic field structures
 that expand out into interplanetary space. They appear like
 X-ray darker areas in the corona of our Sun, but they show a non-zero emission and display a typical temperature  $T_{\rm x,CH} \approx 1$\,MK \citep{Cranmer09.0}.  \citet{Schmitt_97} and \citet{Schmitt12.0},  chasing for the maximum and minimum value of the surface X-ray flux for solar-like stars, suggested that the expected minimum surface flux should occur if the stellar corona is completely covered by CHs, while the maximum if it is covered by active regions. 
In this article we focus on the X-ray properties of the two components of the ultra-low activity binary star GJ\,745\,AB in the context of the $10$\,pc sample of M\,dwarfs. In Sect.~\ref{sect:sample} we describe the new catalog of stars within $10$\,pc by \cite{Reyle_2021} and how we extracted our volume-complete M dwarf sample from that work. In Sect.~\ref{sect:xray} we present our X-ray database and analysis. In Sect.~\ref{sect:discussion}, we discuss the stars of our catalog that have a coronal surface flux so low that we could imagine their corona completely covered by coronal holes.

\section{The M10pc-Gaia sample}
\label{sect:sample} 

 We constructed a volume-complete sample of M\,dwarfs  with spectral type  (SpT)  from M0 to M4 based on the $10$\,pc-catalog published by \citet{Reyle_2021}. We restricted the sample to early M-type dwarfs because we aim  at a sensitive X-ray census of the volume-limited sample. Since X-ray emission drops with later SpT most M\,dwarfs beyond M4 - even those at very close distance - are still  out-of-reach for a systematic survey with current X-ray instrumentation that would require prohibitively long exposure time.

\begin{figure}
    \centering
     \includegraphics[width=0.45\textwidth]{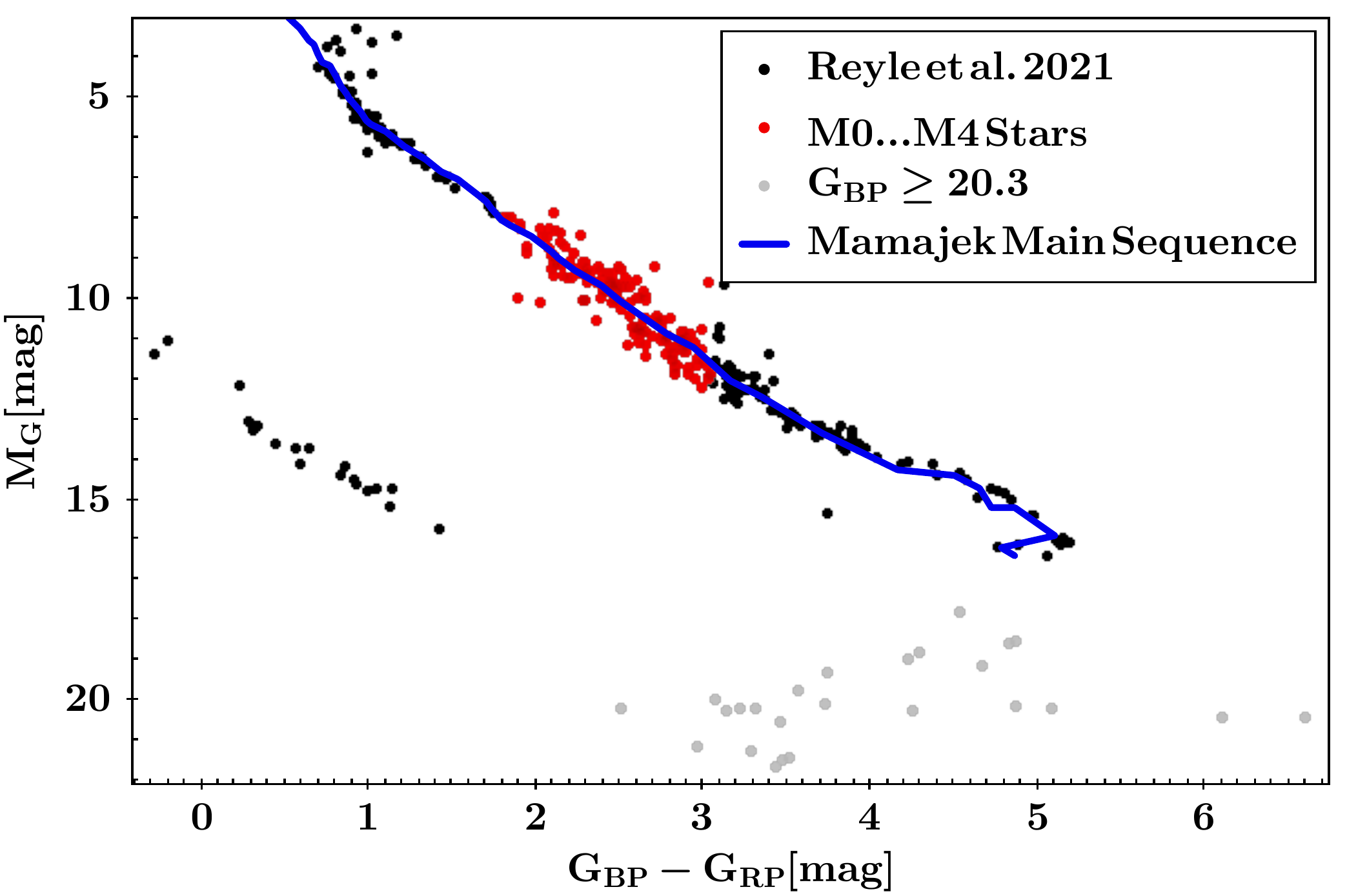}
    \caption{{\it Gaia}  color-magnitude diagram for the $10$\,pc census from \cite{Reyle_2021}. The stars with unreliable $G_{\rm BP}$ magnitude are shown in gray and the selected {\sc M10pc-Gaia sample}  (restricted to SpT M0...M4) is highlighted in red. Superposed is the main-sequence from E.Mamajek's table \footref{note_mamajek}.}
    \label{fig:gaia_cmd}
\end{figure}

The \citet{Reyle_2021} catalog includes $560$ objects that were extracted from the SIMBAD database\footnote{\label{note_symbad} \texttt{\url{http://SIMBAD.u-strasbg.fr}}} \citep{Wenger_2000} with the criterion of 
having a parallax larger than $100$\,mas. Among these, $346$ objects have {\it Gaia} photometry.
In Fig.~\ref{fig:gaia_cmd} we visualize this sample in the {\it Gaia} color-magnitude diagram. 

In the first phase of the sample downselection, we discarded all stars with {\it Gaia} $G_{\rm BP} \geq 20.3$\,mag since for these faint objects the flux in {\it Gaia}'s blue photometer is overestimated leading to an unphysical turnaround of the lower main-sequence that is seen in Fig.~\ref{fig:gaia_cmd} \citep [see discussion in][]{Riello_2021}. 
We then selected all stars that have $G_{\rm BP} - G_{\rm RP}$  color corresponding to M0$-$M4 SpT. 
We performed the association of {\it Gaia} color with SpT using the table from \citet{Pecaut_2013}, integrated with {\it Gaia} photometry and maintained by Eric Mamajek\footnote{see \label{note_mamajek} \texttt{ \url{https://www.pas.rochester.edu/\~ emamajek/EEM\_dwarf\_UBVIJHK\_colors\_Teff.txt}}}.

The sample selected that way consists of $150$ stars. This number is in reasonable  agreement with the extrapolation from the sample  presented by \cite{Schmitt04.0} for a volume of  $6$\,pc radius. These authors counted $37$ stars of spectral type M0...M4, while our {\sc M10pc-Gaia sample} counts $43$ stars within a $6$\,pc distance.
This difference is probably to be attributed to the different ways in which the databases were  collected,  spectral types assigned, and multiplicity treated. \cite{Reyle_2021} consider their catalog complete down to SpT Y2. In fact, as can be seen from Fig.~\ref{fig:gaia_cmd}, the faintest stars  in the {\sc M10pc-Gaia sample} have $G \approx 13$\,mag, many orders of magnitude above the {\it Gaia} sensitivity limit.

Knowledge of the fundamental parameters of the stars is essential for the interpretation of the X-ray data. In particular, the stellar radius ($R_*$) is 
required to determine the X-ray surface flux, 
which is the most useful  parameter to compare the activity levels for a range of stars. 
We calculated $R_*$ 
from $K_{\rm S}$ magnitudes reported in \citet{Reyle_2021}
adopting  
the empirical relation by \cite{Mann15.0}. 

Another essential parameter to achieve a better understanding of 
our M dwarf sample are stellar metallicities. Therefore, we collected metallicity values ${\rm [Fe/H]}$ from the literature for the 
{\sc M10pc-Gaia} stars. \citet{2021A&A...656A.162M} reported metallicities for a sample of 343 M\,dwarfs observed with CARMENES of which 76 stars are in common with the {\sc M10pc-Gaia sample}. GJ\,745\,AB, the binary system on which we will focus in the following, belongs to this group.
 Here we used their metallicity values obtained in the {\sc SteParSyn} run where all parameters were allowed to vary. Measurements of ${\rm [Fe/H]}$ for the remaining stars were collected from \citet{2020ApJ...892...31B,2019A&A...624A..94M,2019ApJ...871...63M,2016ApJ...826..171G,2014AJ....147...20N,2013A&A...551A..36N} and \citet{2012ApJ...748...93R}. Finally, we adopted the metallicity from the Tycho-2 catalog \citep{2006ApJ...638.1004A} for one additional binary pair. In total, we found [Fe/H] values for $135$ of the $150$ stars.

\section{X-ray data base and analysis}
\label{sect:xray}

We compiled an X-ray catalog for the {\sc M10pc-Gaia sample} using both archival data and observations obtained by us through dedicated {\it XMM-Newton} pointings with the purpose of completing the X-ray census of 
early-M\,dwarfs within $10$\,pc of the Sun.
The archival data comes from ROSAT/PSPC \citep{Briel86.0} observations both during the all-sky survey phase and the subsequent pointed phase, from the  all-sky survey 
of the {\it extended ROentgen Survey with an Imaging Telescope Array} (eROSITA; \citep{Predehl21.0}) on the Spectrum-Roentgen-Gamma (SRG) mission 
and from pointed {\it XMM-Newton} observations as explained in Sects.~\ref{subsect:xray_RASS} to~\ref{subsect:xray_xmm}. Targets that at the beginning of the project had no X-ray detection  in any of these databases are  observed in the framework of an {\it XMM-Newton} {\sc fulfil} program (PI Stelzer Obs ID \,084084, 086030). 

The full X-ray catalog will be published after the X-ray census is completed. 
To date, we still need to observe $9$ out of the $150$ M\,dwarfs within $10$\,pc of the Sun. 
These are part of an {\it  XMM-Newton}/AO22 campaign (PI Stelzer; ObsID 092126) in continuation of the above mentioned {\sc fulfil} program that started in AO18. Here we describe how we compiled the X-ray data base and how we extracted the basic parameter, a homogeneously determined  X-ray flux for all sample stars.

The most delicate step in the construction of a homogeneous X-ray catalog is the calculation of the flux from the count rate for observations acquired with instruments that cover  different energy bands, here {\it XMM-Newton}, eROSITA and ROSAT. 
We, therefore, computed separately for the observations of each data base X-ray fluxes with an instrument-specific rate-to-flux conversion factor ($CF$).
All fluxes are calculated for the $0.1-2.4$\,keV ROSAT energy band which is the band that is most widespread in the literature. Also, this energy band has been used in the construction of empirical relations between X-ray and EUV flux/luminosity \citep[e.g.][]{SanzForcada11.0, Chadney15.0}, and thus allows for the most direct conversion between the two energy bands. We note that to cover the ROSAT band we had to  slightly extrapolate the  energy range at the low end covered by the other X-ray instruments, by $0.1$\,keV for eROSITA and {\it XMM-Newton}. Since our flux calculation is based on a given spectral model this means that we implicitly assume that at the low-energy end no additional spectral component  contributes.
In Table~\ref{cf_table} we list the $CF$ that we used for the three instruments to calculate a homogeneous X-ray flux in the ROSAT band. Explanations on how we derived these values are found in the remainder of this section.
\begin{table}
\caption{Conversion factor between count rate measured with different X-ray instruments and flux in the  $0.1-2.4$\,keV ROSAT band.
}\label{cf_table}
\centering
\begin{tabular}{lr}
\hline \hline
Instrument & $CF$ [${\rm erg/cm^2 /cts}$] \\ \hline
ROSAT & $5.77 \cdot 10^{-12}$  \\
eROSITA & $8.78 \cdot 10^{-13} $  \\
{\it XMM-Newton} & $1.38\cdot 10^{-12} $ \\
\hline
\end{tabular}
\end{table}

\subsection{ROSAT All-Sky Survey}
\label{subsect:xray_RASS}

We extracted data from the {\it Second ROSAT All-Sky Survey Point Source Catalog}  \citep[2\,RXS;][]{Boller16.0}. 2\,RXS is a revised version of the earlier bright \citep{Voges99.0} and faint \citep{Voges00.0} source catalogs. It tabulates count rates in the $0.1-2.4$\,keV band for more than $130 000$ X-ray sources distributed over the whole sky.

To match the {\sc M10pc-Gaia sample} with 2\,RXS we followed  the procedure of \cite{Stelzer13.0}. We  first extrapolated the star coordinates reported by \citet{Reyle_2021} back to Oct 1, 1990, a date representative for the observing epoch of the RASS which lasted from Aug - Dec 1990. Then, we performed a cross-match with a radius of $40^{\prime\prime}$ \citep{Neuhaeuser_95} , obtaining that $86$
stars of our sample were detected in the RASS.

For the conversion of the 2\,RXS count rates to fluxes we used  as the $0.1-2.4$\,keV conversion factor the value of $CF_{\rm ROSAT} = 5.77 \cdot 10^{-12}\,{\rm erg/cm^2 /cts}$ that was determined by \cite{Magaudda20.0} with the Mission Count Rate Simulator {\texttt{ WebPIMMS}}\footnote{\texttt{\url{https://heasarc.gsfc.nasa.gov/cgi-bin/Tools/w3pimms/w3pimms.pl}}} for a 1T-{\sc APEC} model with a temperature of $kT = 0.5$\,keV.
This spectral model is the average of the best fit  parameters obtained for {\it XMM-Newton} and {\it Chandra} spectra of  \cite{Magaudda20.0} M dwarf sample.

\subsection{eROSITA All-Sky Survey}
\label{subsect:xray_eRASS}

 We have used the merged catalog of the first four eROSITA all-sky surveys (eRASS:4) in the version available to the eROSITA\_DE consortium in October 2022\footnote{The consortium-internal file name of the eRASS catalog we used is  all\_s4\_SourceCat1B\_221031\_poscorr\_mpe\_photom.fits.}. 
 The eROSITA catalogs available to us have been produced at Max-Planck Institut f\"ur extraterrestrische Physik and they comprise data from the western galactic hemisphere ($l \ge\,180$°) which is the half of the sky with german data rights. Before the match with a radius of $30^{\prime\prime}$ 
we have translated the coordinates of the {\sc M10pc-Gaia} stars to the mean observing date of the four eRASS surveys, 15 December 2020. 
None of the sample stars has a proper motion as high as to be able to move outside the $30^{\prime\prime}$  match radius within one year, namely the time difference between the above mentioned mean eRASS date and the beginning of the first and the end of the forth survey. With this catalog match we found $69$ eRASS detections among the {\sc M10pc-Gaia} stars. 

The eRASS:4 catalog holds count rates in a single energy band ranging from $0.2-2.3$\,keV. We converted them into flux in the ROSAT band by means of the conversion factor $CF_{\rm eRASS} = 8.78 \cdot 10^{-13}\,{\rm erg/cm^2 /cts}$. 
We derived this value
from a combination of one and two temperatures APEC models for a sample of early M dwarf stars studied in \cite{Magaudda22.0}. The $CF_{\rm eRASS}$ of this work appears slightly higher than the value published in \cite{Magaudda22.0} because they adopted the eROSITA broad band (0.2-5.0 keV), while we limited our analysis to the energy band of ROSAT (0.1-2.4 keV).

\subsection{ROSAT pointed catalog, 2RXP}
\label{subsect:xray_2RXP}

For the search of {\sc M10pc-Gaia} stars in pointed ROSAT observations, we cross-matched their coordinates with those in the  {\it Second ROSAT Source Catalog of Pointed Observations} \citep{Rosat_2000} 
which provides count rates in the $0.1-2.4$\,keV band from stars detected in pointed ROSAT/PSPC observations.

We proceeded with the 
propagation of the coordinates to the epoch of the X-ray data analogous to the steps discussed above for 2\,RXS and eRASS:4. The difference is that the 2\,RXP data is spread over a much longer time range (about seven years). Therefore, a mean proper motion correction may not yield a position at the time of the ROSAT observation that is accurate enough to retrieve the X-ray 
counterparts for all stars.
Instead, individual proper motion corrections must be applied for each target to its ROSAT observing date. Since we did not know this date a priori, we started with an initial very large match radius of $260^{\prime\prime}$. This value is 
motivated by the maximum proper motion of our stars between 
the year 1991, namely  the epoch of the first pointed ROSAT observations, and their {\it Gaia} position.  

The same value of  $CF_{\rm ROSAT} = 5.77 \cdot 10^{-12}\,{\rm erg/cm^2/cts}$ as for the RASS detections (see Sect.~\ref{subsect:xray_RASS}) was used to obtain the $0.1-2.4$\,keV fluxes from the count rates listed in 2\,RXP.

\subsection{ \it XMM-Newton}
\label{subsect:xray_xmm}

 To find targets detected with {\it XMM-Newton} we have searched the 4XMM-DR11 catalog \citep{Webb20.0} and, for more recent observations, we consulted directly the {\it XMM-Newton} archive\footnote{\url{http://nxsa.esac.esa.int/nxsa-web/##search}}. For simplicity we made use only of the most sensitive of the EPIC instruments, the pn CCDs.

 Similar to the case of the pointed ROSAT catalog, since there are up to $16$\,yrs between the epoch of the {\it XMM-Newton} observations in 4XMM-DR11 and the {\it Gaia}-eDR3 epoch, we had to perform individual proper motion corrections for our target stars.
The initial match radius we used was $180^{\prime\prime}$, motivated by the maximum proper motion expected for any of our targets within the maximum possible time difference of $16$\,yrs. 
The coordinates of the stars that present one or more {\it XMM-Newton}  
counterparts
for this large search radius were then propagated to the epoch of the specific {\it XMM-Newton}  observation. The subsequent refined cross-match was performed with a radius of $15^{\prime\prime}$. 
This way, we found that $28$ stars of the {\sc M10pc-Gaia sample}
have at least one  detection in the 4XMM-DR11 catalog.

The {\it XMM-Newton} archive holds more recent observations than 4XMM-DR11 for an additional eight  {\sc M10pc-Gaia} stars from our dedicated {\it XMM-Newton} {\sc fulfill}  survey.
For these observations we used the standard  \texttt{SAS} pipeline for the source detection and the determination of the count rate. X-ray counterparts were then identified by means of a cross-match, with a radius of  $15^{\prime\prime}$, between the position of the detected X-ray sources and those of the {\it Gaia} 
position propagated to the date of the {\it XMM-Newton} observation.

One star required a special treatment for its detection: GJ\,643 was observed with {\it XMM-Newton}, but it is located in the wings of the EPIC/pn Point Spread Function (PSF) of the much brighter star GJ\,644 which is also a member of the {\sc M10pc-Gaia sample}. With a special background subtraction and thanks to a flare-event on GJ\,643 we managed to detect the star in spite of the fact that the noise by far exceeds the signal. The  analysis of the EPIC/pn data for this object is detailed in Appendix~\ref{app:gj643}. 

For all {\sc M10pc-Gaia} stars  observed with {\it XMM-Newton}, whether extracted from the 4XMM-DR11 catalog or directly from the archive, we followed \citet{Magaudda20.0} and converted 
their EPIC/pn count rates in the full $0.2-12$\,keV energy band into flux in the ROSAT band with the conversion factor value of $CF_{\rm XMM} = 1.38\cdot 10^{-12}\,{\rm erg/cm^2 /cts}$,
calculated with {\texttt{ WebPIMMS}} for a 1T-{\sc APEC} model with a temperature of $kT = 0.5$\,keV.

\subsection{Summary on X-ray detections}
\label{subsect:xray_det_summary}

\begin{figure}
    \centering
    \includegraphics[width=0.5\textwidth]{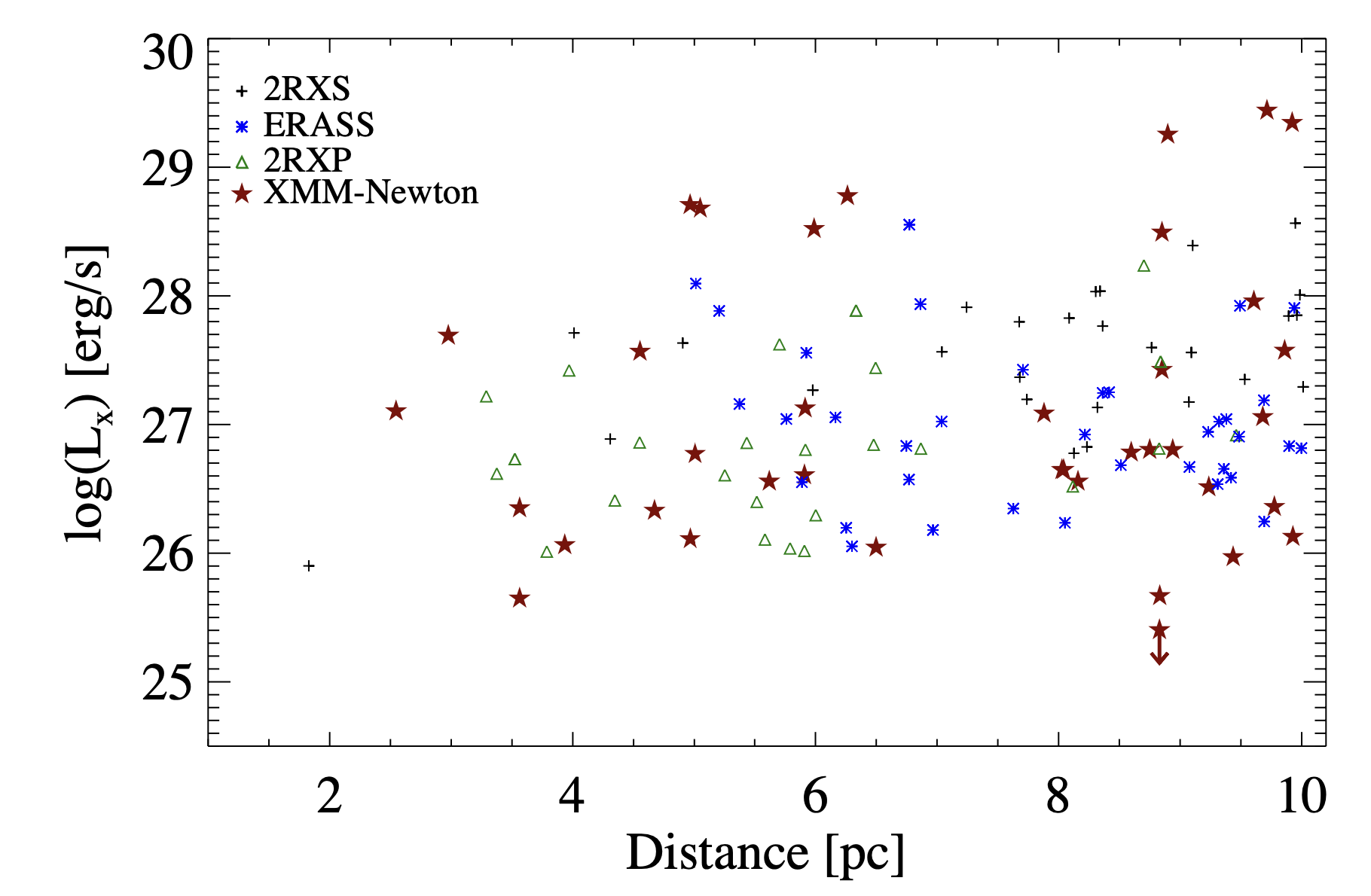}
    \caption{X-ray luminosity versus distance for the {\sc M10pc-Gaia sample}. For stars detected more than once the $L_{\rm x}$ value from the deepest survey was used in the order explained in the text. Among all other stars only one has remained undetected, GJ\,745\,A, which has an upper limit indicated by the downward pointing arrow. 
}
    \label{fig:lx_dist_surveytype}
\end{figure}

Many stars have more than one X-ray detection which allows for a variability study that we defer to a future work. For this article, we focused on the overall range of X-ray emission levels of early-M\,dwarfs.
Fig.~\ref{fig:lx_dist_surveytype} shows the X-ray luminosities for the whole {\sc M10pc-Gaia} sample. Since some stars have been observed more than once, we plot in this figure the X-ray detection from the overall deepest data, 
namely {\it XMM-Newton}, 2RXP, eRASS and RASS, in this order. Specifically, Fig. \ref{fig:lx_dist_surveytype} holds 37 stars detected with  {\it XMM-Newton}, 28 from 2RXP, an additional 38 stars from eRASS:4 and 30 from 2RXS.

Besides the nine  stars that have not yet been observed, and that are part of our ongoing  observational campaign, there is  just one object, GJ\,745\,A, that has deep X-ray data from our dedicated {\it XMM-Newton}  survey, and yet  was not detected. 
In Sect.~\ref{subsect:xray_gj745a} we explain how we extracted the upper limit to its X-ray emission.

\subsection{A very deep X-ray upper limit for GJ\,745\,A}
\label{subsect:xray_gj745a}

GJ\,745\,A is part of a multiple system together with GJ\,745\,B, its proper motion companion.
The system
was observed twice with {\it XMM-Newton}. The first observation was taken on Sept 27-28 2019 (Obs ID: 0840843401) for $32.4$\,ks, with
GJ\,745\,B as main target. One year later, on Sept 20-21 2020, GJ\,745\,A was observed for $31.7$\,ks in a dedicated observation (Obs ID: 0860303001).
With a  separation of $114.06^{\prime\prime}$ \citep{Andrews17.0} the two components of GJ\,745 are clearly resolved with {\it XMM-Newton}. 

We analyzed both {\it XMM-Newton} EPIC/pn observations using the Science Analysis Software (\texttt{SAS}) version 19.1.0 developed for the satellite. By examining the high energy events ($\geq 10$\,keV) across the full EPIC/pn detector, we excluded
the time intervals affected by solar particle background. We note that the latest observation (0860303001) is affected from a large background event, that reduced the good time intervals and consequently the exposure time by about 50\%.
We filtered the data for pixel patterns ($0 \le$ pattern $\le 12$), quality flag (flag = 0) and events channels ($PI \ge 150$). The source detection was performed in three energy bands: $0.2 - 0.5$\,keV (S), $0.5 - 1.0$\,keV (M), and $1.0 - 2.0$\,keV (H). 

GJ\,745\,A was not detected in either of the two observations. Making use of the \texttt{SAS} tool \texttt{ ESENSMAP}, we calculated the sensitivity map in the $0.2-12.0$\,keV band and derived the two upper limits at the {\it Gaia} position of the star
propagated to the date of each observation.
The two values are reported in Table~\ref{upper_limit_table}.
\begin{table}
\caption{Upper-limit count rates and effective exposure time after removal high-background time intervals for GJ\,745\,A during the two {\it XMM-Newton} observations.
}\label{upper_limit_table}
\centering
\begin{tabular}{cccc}
\hline \hline
Obs.ID & Obs.Date & u. l. rate [cts/s] & Exp. Time [s] \\ \hline
$0840843401$ & 2019-09-27 & $0.00197$ & $16272 $  \\
$0860303001$ & 2020-09-20 & $0.00272$ & $14452 $ \\
\hline
\end{tabular}
\end{table}
The difference of the exposure time and the different level of background of the two observations  leads to 
different upper limit count rates. In the following we use the lower, that is more sensitive,  value ($0.0020$\,cts/s) as upper limit  count rate of GJ\,745\,A.

Converting the count rate to luminosity in the $0.1-2.4$\,keV band, as explained in Sect.~\ref{subsect:xray_xmm}, 
 we obtained $\log{L_{\rm x}}\,{\rm[erg/s] } < 25.4$. For the star's stellar radius of $0.34 \,{\rm R_{\odot}}$ the  X-ray surface flux is $\log{F_{\rm X,SURF} < 3.55}\, {\rm erg\ cm^{-2}\ s^{-1}}$.
 Here we have used the conversion factor that applies for a coronal temperature of $0.5$\,keV. In Sect.~\ref{sect:discussion} we show that the faint X-ray emission level of GJ\,745\,A is consistent with the emission of solar coronal holes, which have lower temperature. If we 
assume that the temperature in GJ\,745\,A's corona is $0.1$\,keV, 
the resulting X-ray surface flux
would be $\log{F_{\rm X,SURF} <  3.74}\, {\rm erg\ cm^{-2}\ s^{-1}}$. 

\section{Discussion}\label{sect:discussion} 

\begin{figure}
    \centering
    \includegraphics[width=0.5\textwidth]{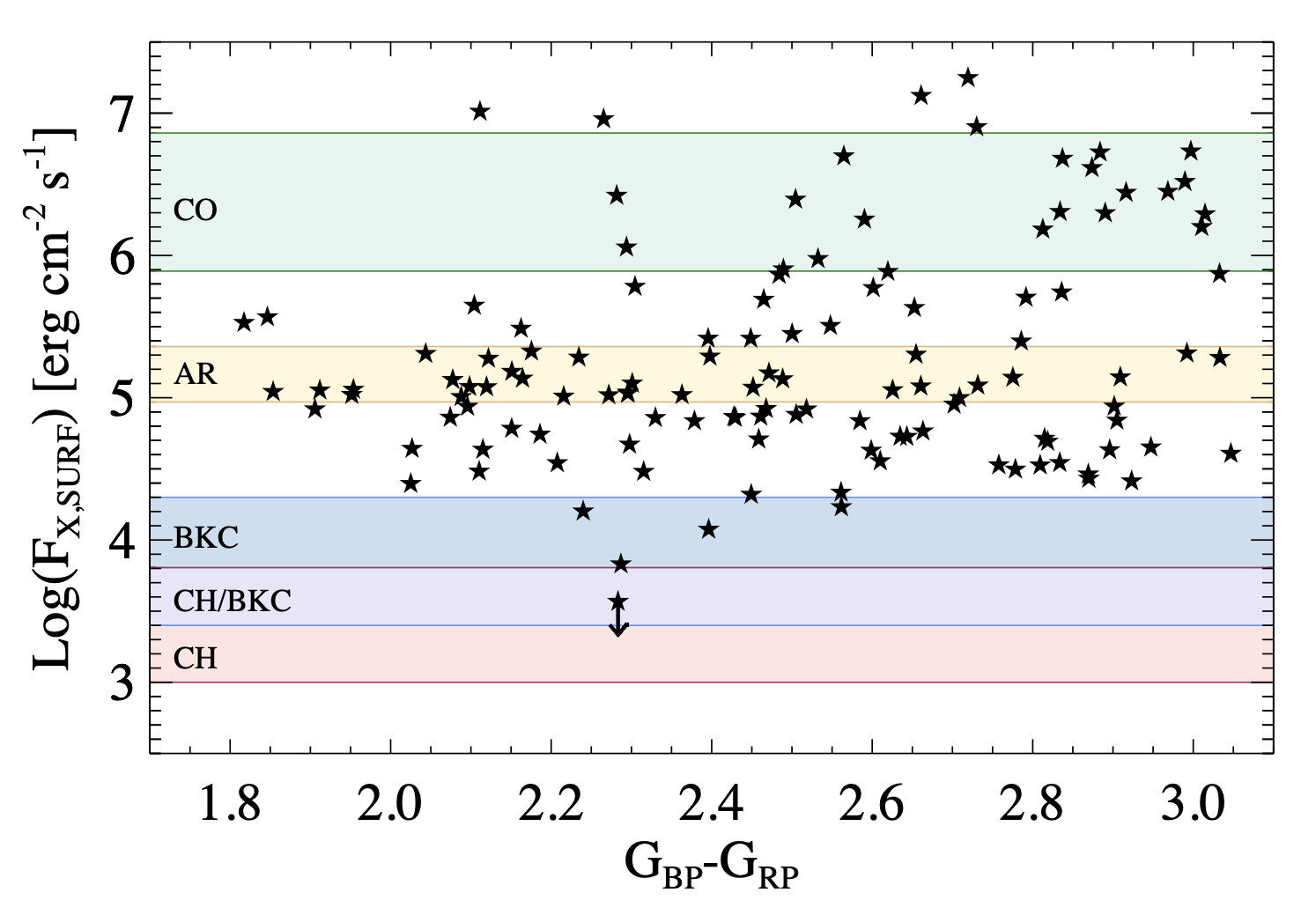}
    \caption{X-ray surface flux versus {\it Gaia} $G_{\rm BP} - G_{\rm RP}$ color  for the {\sc M10pc-Gaia  Sample}.  Each star is represented by its mean X-ray emission level from our 
    multi-mission data base. Colored areas mark the range of surface flux
    for four typical  non-flaring magnetic regions in the solar corona, coronal holes (CH), the background corona (BKC), the active regions (AR) and the cores of active regions (CO); see Table \ref{tab:solar_regions}. 
    The flux range of BKC and CH overlap. 
    Only two stars are located within or in the immediate vicinity of the `coronal hole' area, the binary pair GJ\,745\,A and~B, where the upper limit is for the A component of the system.
 }
    \label{fig:Fxsurf_dist}
\end{figure}

Fig.~\ref{fig:lx_dist_surveytype} shows a remarkable distribution of the $L_{\rm x}$ values
in the {\sc M10pc-Gaia Sample}. Next to a scatter of the detections by about two orders of magnitude a lower envelope is seen that is located at  approximately $\log{L_{\rm x}}\,{\rm  [erg/s]} = 26$. 
This value shows no obvious dependence on distance, indicating that there is no sensitivity-related X-ray detection bias. Leaving apart the $9$ stars that still need to be observed in our XMM-Newton program, the {\sc M10pc-Gaia sample} is the first truly volume-complete M dwarf sample in both the optical and in X-rays. 

The most universal diagnostic for coronal brightness is the surface X-ray flux, $F_{\rm X,SURF}$, since it is independent of the stellar radius. In Fig.~\ref{fig:Fxsurf_dist} we show $F_{\rm X,SURF}$ versus {\it Gaia} color for the {\sc M10pc-Gaia sample}. In the solar corona, individual types of magnetic structures characterized by   different X-ray brightness can be distinguished in images, and their emitted flux quantified. Such studies were carried out in the project {\it The Sun as an X-ray star}, see e.g. \cite{Orlando01.0}. From a comparison of the observed range in the X-ray emission level of a stellar sample with these solar structures it is, therefore, possible to estimate which types of magnetically active regions dominate in  the coronae of the  stars.

\subsection{The X-ray emission from solar coronal  structures}\label{subsect:discussion_CH}
Within {\it The Sun as an X-ray star project}, {\it Yohkoh} observations have been used to quantify the emission measure distributions of different types of magnetic structures in the solar corona \citep{Orlando00.0, Orlando01.0,Peres_00, Reale_01}. These structures are defined by their surface brightness, in increasing order from background corona (BKC), over active regions (AR) to cores of active regions (CO); see \cite{Orlando01.0}. The solar {\it Yohkoh} emission measure distributions for each of these types of regions in the corona of the Sun can be converted into a synthetic X-ray spectrum considering the instrumental response of the facility in question. Since we are using the ROSAT band, the appropriate instrument is the ROSAT/PSPC. Once the synthetic X-ray spectrum has been obtained its flux can be calculated for a selected energy band. This way, the typical surface flux of solar BKC, AR and CO can be determined. 
For our purpose {\it Yohkoh} data from July 1996 was used when there was only one active region (emerged on July 4th) on the Sun \citep[see][]{Orlando_04},
and the range of X-ray fluxes for the different types of regions derived from these data are listed in Table~\ref{tab:solar_regions}. In Fig.~\ref{fig:Fxsurf_dist} we overlay the range of BKC, AR and CO as colored areas.
 Clearly, it cannot be excluded that on some other date the Sun displayed regions with slightly  different properties, but the ranges given in Table~\ref{tab:solar_regions} should be approximately representative for each type of magnetic solar  structure.   
 
The lowest expected X-ray emission level of a late-type star is the one 
of a corona characterized by open field lines where hot plasma escapes into space. On the Sun, this configuration 
corresponds to the so-called coronal holes.   The surface flux of (solar) coronal holes in the $0.1-2.4$\,keV ROSAT band has been computed by \cite{Schmitt12.0} for a range of temperatures observed in coronal holes on the Sun. 
In Fig.~\ref{fig:Fxsurf_dist} we include the range of X-ray surface flux values for a solar coronal hole 
corresponding to  plasma temperatures from $1$\,MK (upper bound of the colored stripe) up to  $2$\,MK (lower bound of the stripe). The CH flux corresponds, as predicted, to the lowest X-ray emission levels observed on the Sun, but the CH region shows some overlap with the BKC.  
The BKC corresponds to faint and diffuse regions with pixels with good S/N \citep[i.e. with more than $10$  photons per pixel; see Fig.~12 in][]{Orlando00.1} and surface intensity below a threshold to exclude ARs and COs \citep[see Fig.~2 in][]{Orlando01.0}. This selection criterion effectively eliminates regions with
very low emission (i.e. pixels with less than 10 photons), resulting in the exclusion of the weakest parts of CHs. Consequently, there is a significant difference in the flux level between BKC and CHs.

\begin{table}[]
    \centering
        \caption{X-ray surface fluxes 
        for  magnetic structures (BKC, AR and CO) in the corona of the Sun obtained from {\it Yohkoh} data collected in July 1996 (during a minimum in the solar cycle) within  the {\it Sun as an X-ray star} framework. Values are logarithmic and  given in units of ${\rm erg/cm^2/s}$. The fluxes  of solar 
        CHs for a range of plasma temperatures between $1$\,MK (maximum)  and $2$\,MK (minimum) are from  \citep{Schmitt12.0}. All values refer to the ROSAT energy band.
        }
    \label{tab:solar_regions}
    \begin{tabular}{lrrrr}\hline
    Region & minimum & maximum & median & average \\ \hline
    CH  &      3.00 &      3.78   &   ...     &  ... \\
  BKC     &  3.37       & 4.30     & 4.14     &  4.10 \\
AR &       4.97      & 5.36    &  5.10    &   5.10 \\
CO  &      5.89 &      6.86   &   6.19     &  6.23 \\
\hline
    \end{tabular}
\end{table}

\begin{table*}[]
    \centering
     \caption{Activity-related properties of CH M\,dwarfs in the {\sc M10pc-Gaia sample}. For GJ\,745B the X-ray parameters represent the time-averaged values.}
    \begin{tabular}{lrrrrrrcc} \hline
        Name & $\log{F_{\rm X,SURF}}$\, $[{\rm erg/cm^2 /s}]$ & $\log{L_{\rm x}} $ \, $[{\rm erg/s}]$ & [Fe/H] [dex]& $u$ [km/s] & $v$ [km/s] & $w$  [km/s] & EW\,H$\alpha$ [\AA] \\ \hline
        GJ\,745\,A & $< 3.55$ & $< 25.40$ & $-0.68$ & $-45.0$ & $+23.4$ & $+21.6$ & $-0.108$ \\ 
        GJ\,745\,B & $3.82$ & 25.67 & $-0.67$ & $-44.4$ & $+23.6$ & $+22.0$ & $-0.104$ \\
        \hline
    \end{tabular}
    \label{tab:CHstars}
\end{table*}

\subsection{Coronal hole-like M\,dwarfs}\label{subsect:discussion_CHstars_in_context}
Several nearby M\,dwarfs appear to be very X-ray quiet as inferred from the comparison with the solar data in Fig.~\ref{fig:Fxsurf_dist}. 
However, only two stars have X-ray surface fluxes 
as low as solar CHs.
This is the binary pair GJ\,745\,A and~B, with the primary component being the only star in our sample that is not detected. If the lowest X-ray emission level of M\,dwarfs is represented by the flux of a solar CH, these stars might  thus be completely covered with CHs. Since their flux lies in the overlapping locus between CH and BKC, their emission could also be explained with a quiet corona without active regions. A more realistic scenario may correspond to a combination between CH and BKC, although
a small contribution by brighter magnetic structures (AR and CO) cannot be excluded. Analogously, 
for the more active stars it is not possible to clearly associate their X-ray emission to one type of magnetic structure since any real stellar corona, just as our Sun, can be expected to be composed of a  mixture of CH, BKC, AR and
CO (with, occasionally, flaring structures), and the relative covering fraction for each of these structures cannot be quantified with a simple flux measurement. 

Although the X-ray emission of GJ\,745 can be explained by either CH or BKC alone  
they are the only stars
of our sample 
consistent with
a CH-like corona.  Therefore we focus in the following on a description where the corona is a combination of
some area fraction of BKC, AR or CO with the remaining larger part covered with CHs, that is
\begin{equation}
L_{\rm x,*} / (4 \pi R_*^2) = F_{\rm surf,REG} \cdot f + F_{\rm surf,CH} \cdot (1-f).
\label{eq:ff}
\end{equation}
Here, REG stands for one of the solar magnetic regions (BKC, AR or CO) and $f$ is the filling factor that is a percentage area coverage. 

The two components of the GJ\,745 binary are twins, with equal radii ($R_* = 0.34\,{\rm R_\odot}$) and masses ($M_* = 0.34\,{\rm M_\odot}$). The apparently X-ray dark star, GJ\,745\,A, has an upper limit of 
$\log{F_{\rm X,SURF,GJ745A}}\,{\rm [erg/cm^2/s]} <3.55$, and 
its companion, GJ\,745\,B, is a very weak X-ray source with 
$\log{F_{\rm X,SURF,GJ745B}}{[ \bf \rm erg/cm^2/s]} = 3.82$ (see Table~\ref{tab:CHstars}). 
For our evaluation of Eq.~\ref{eq:ff} we consider for CHs the minimum flux (lower bound in Fig.~\ref{fig:Fxsurf_dist}) of $\langle F_{\rm CH\,}{\bf \rm erg/cm^2/s} \rangle = 10^{3.0}$ and for REG we used the minimum values from Table~\ref{tab:solar_regions}. This way we obtain an upper limit to the flux contribution from the brighter region, REG. 
With this approach, the 
observed X-ray 
upper limit for GJ\,745\,A 
can be explained 
with a corona covered in large part with CHs except for
$2.7$\,\% of AR or $0.3$\,\% CO.
On the other hand, assuming the star to be completely covered with BKC at the minimum solar flux value, the resulting luminosity would be below the observed X-ray upper limit, and hence a BKC-like corona is also compatible with the observation of GJ\,745\,A.  
However, if we replace the minimum solar BKC flux by the average, the scenario 
 is that
of a star with dominating emission of CH and a BKC filling factor area of $21.8$\,\%.  

The  time-averaged 
X-ray surface flux of GJ\,745\,B, like that of its companion, is compatible with a scenario where the emission is entirely explained by BKC. A combination of CH+BKC, both at the minimum of the solar values is not able to explain the observed flux of GJ\,745\,B. If, instead, we consider the star to be dominated by CH and BKC at the mean solar flux, we obtain a BKC filling factor of $48$\,\%. Another scenario would be that of
$6.0$\,\% of the corona covered with solar-like AR and the rest with CH.
Replacing the AR with CO we find for GJ\,745\,B a maximum filling factor with CO of $0.72$\,\%. As we show below in Sect.~\ref{subsubsect:discussion_CHstars_in_context_flare}, GJ\,745\,B underwent a flare during the {\it XMM-Newton} observation. Since flares on the Sun take place in the COs of ARs the presence of an erupting CO on the corona of GJ\,745\,B is plausible. 

\subsubsection{Coronal brightness and  temperature}
\label{subsubsect:discussion_CHstars_in_context_Tx}

\begin{figure}
    \centering
    \includegraphics[width=0.5\textwidth]{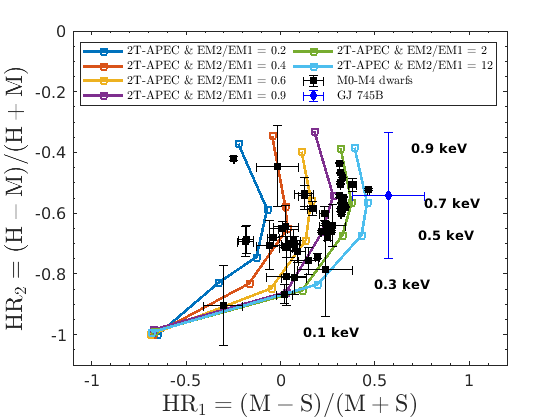}
    \caption{Hardness ratios of the stars from the {\sc M10pc-Gaia sample} that are detected with {\it XMM-Newton} (filled black squares), with  the CH star (in blue) 
    highlighted. Superposed on the data is a grid of 2T-APEC models with a fixed $kT_{\rm 1}$ at $0.1$\,keV, $kT_{\rm 2}$ and $EM2/EM1$ ranging from $0.1$\,keV to $0.9$\,keV and $0.2$ to $12$, respectively (see Sect.~\ref{subsubsect:discussion_CHstars_in_context_Tx} for more details).}
    \label{fig:HR1_HR2}
\end{figure}

For GJ\,745\,A the upper limit does not provide us any information on its coronal temperature. However its companion GJ\,745\,B was detected with $49 \pm 9$ EPIC/pn counts from ObsID 0840843401. The collected counts are too few for a spectral fit, therefore we studied its spectral shape through hardness ratios. 

For the hardness ratio analysis we use standard {\it XMM-Newton} energy bands, $0.2-0.5$\,keV, $0.5-1.0$\,keV and $1.0-2.0$\,keV, henceforth named S, M and H for the soft, medium and hard part of the X-ray spectrum.
The hardness ratios are  defined as
$HR_{\rm 1} = (Rate_{\rm M}-Rate_{\rm S})/(Rate_{\rm M}+Rate_{\rm S})$
and 
$HR_{\rm 2} = (Rate_{\rm H}-Rate_{\rm M})/(Rate_{\rm H}+Rate_{\rm M})$, 
where $Rate$ is the net source count rate resulting from the source detection process
in the respective energy band.  

Fig.~\ref{fig:HR1_HR2} shows the hardness ratio plot for all {\sc M10pc-Gaia sample} stars observed and detected with {\it XMM-Newton}, including the CH star GJ\,745\,B which is highlighted with a blue colored plotting symbol. Superposed on the data is a grid calculated for a 2T-APEC model from a simulated {\it XMM-Newton} spectrum using the EPIC/pn response matrix and the exposure time of the observation of  GJ\,745\,B amounting to $32$\,ks. The low-temperature component is fixed on $kT_{\rm 1} = 0.1$\,keV and the second component, $kT_{\rm 2}$, varies from $0.1$ to $0.9$\,keV in steps of $0.2$\,keV. 
The coronal abundance of both spectral components was set to $0.3$ times the solar value as typical for late-type stars  \citep[e.g.][]{Favata2000,vandenBesselaar2003,Robrade_05,Maggio2007}
using the library from \cite{Aspl2009}.
A third parameter is the ratio of the two emission measures ($EM2/EM1$). In a collisionally ionized plasma the $EM$ 
is a measure for the X-ray emitting power. It is given by the volume integral of the product of the densities of electrons and ions at a given temperature, where the latter ones are related to the hydrogen density through the elemental abundance.
Since $HR$ is a normalized quantity the absolute values for the two $EM$s do not play a role here. We computed our model for values of the ratio of the emission measures varying from $0.2$ to $12.0$, i.e. from where the softer temperature component is responsible for most of the emission to where the harder component dominates.
From the comparison of this model grid with the data we see that all stars have negative $HR2$, indicating an only weak contribution from emission at energies $> 1$\,keV, namely in the band $H$.
This is consistent with the uppermost value for $kT_{\rm 2}$ in the grid of $0.9$\,keV. 
While in our sample there are stars with very soft emission, i.e. $HR_{\rm 1}<0$, for the majority of stars the dominating emission is in band $M$. The empirical upper boundary of our sample in terms of $HR1$ is matched well by the model with $EM2/EM1  =2$, but the stars with the highest values of $HR1$ require $EM2/EM1 = 12$,  corresponding to coronae where the softer component
is of little importance. 
We verified 
the negligible contribution of the soft component
by comparing this model with the hardness ratios from a simulated 1T-APEC spectrum. This latter one, indeed, closely follows our upper boundary 2T-model (with $EM2/EM1 = 12$) which thus represents a point of saturation for our two-temperature grid.  

Remarkably, GJ\,745\,B is located at the upper boundary in terms of $EM2/EM1$, presenting the highest value of $HR_{\rm 1}$ in the whole sample. 
The model that best describes 
the hardness ratios of GJ\,745\,B 
is the one with $kT_{\rm 1} = 0.1$\,keV, $kT_{\rm 2} = 0.7$\,keV and $EM_{\rm 2}/EM_{\rm 1} = 12.0$, yielding an emission measure weighted mean coronal temperature of $0.65$\,keV.

\begin{figure}
    \centering
    \includegraphics[width=0.5\textwidth]{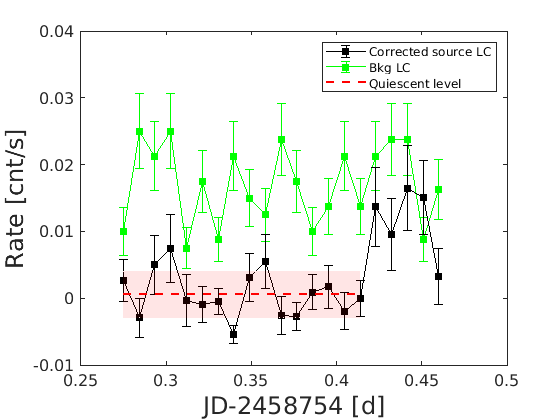}
    \caption{Background-subtracted EPIC/pn light curve of GJ\,745\,B (black) and light curve of the background (green), both with a bin size of $800$\,s;  see  Sect.~\ref{subsubsect:discussion_CHstars_in_context_Tx}.
   We also show the quiescent rate with its standard deviation (red dashed line and shade, respectively) that we adopted for the calculation of the flare energy as explained in Sect.~\ref{subsubsect:discussion_CHstars_in_context_flare}.
  The light curve is shorter than the nominal exposure time due to high and variable background at the end of the observation.}
    \label{fig:GJ745B_lc}
\end{figure}

\subsubsection{A flare on a coronal hole star}\label{subsubsect:discussion_CHstars_in_context_flare}

We have extracted the EPIC/pn light curve of GJ\,745\,B for the same {\it XMM-Newton} energy band used for the hardness ratio ($0.2-2.0$\,keV)   and with a time bin size of $800$\,s (see Fig.~\ref{fig:GJ745B_lc}). The source counts were  extracted from a region of $30^{\prime\prime}$ and the background from a region of $70^{\prime\prime}$ radius.  
After background subtraction the count rate in the light curve is consistent with a non-detection, except for the last hour of the observation in which the rate rises from $\sim 0$\,cnt/s to $\sim 0.015$\,cnt/s. This evident flare is the likely explanation for the high coronal temperature identified in the hardness ratio analysis.

We computed the energy emitted during the flare ($E_{\rm flare}$) adopting as  ``quiescent'' activity level the average count rate before the rise, amounting to \num{5.4e-4}\,cnt/s. This rate is represented by the red dashed line in Fig.~\ref{fig:HR1_HR2} where we also show its standard deviation as red shaded area. We extracted the total flare counts by subtracting the quiescent rate from the rate of the last five bins, multiplying by the time bin size and summing over all elements. This way, we found $\log E_{\rm flare}\,{\rm [erg]} = 29.72\pm0.09$ with the {\it Gaia}-DR2 distance ($8.83\pm0.01$\,pc) and the $CF$ given  in Sect.~\ref{subsect:xray_xmm}.
Although  this $CF$ was computed for a slightly different energy band, it can be applied here because the  time-averaged count rate in the light curve of Fig.~\ref{fig:GJ745B_lc} is consistent with the count rate listed in the 4XMM-DR11 catalog for the $0.2-12$\,keV band, hence there is no significant emission outside the narrower energy range we used for the light curve. 
The flare peak luminosity extracted from the last four  bins with the highest count rate is $L_{\rm x,flare}[\,erg/s] = 26.2\pm 0.1$\,erg/s. The pre-flare quiescent count rate, red dashed line in Fig.~\ref{fig:GJ745B_lc}, is consistent with zero. Therefore, GJ\,745\,B is undetected outside the flare event. We have extracted the upper limit from the  sensitivity map for the out-of-flare time interval, and find $L_{\rm x,quies}[\,erg/s] < 25.52$. 

Outside this flare event the star is undetected, 
which means that its emission state is 
similar to that of its sibling GJ\,745\,A.
At first sight it seems remarkable to observe a flare on a  star that is otherwise X-ray dark at the level of CH-emission.   
However, it is to be kept in mind that the   emission we observe is averaged across all  magnetic structures present on the stellar surface,  and therefore  although most of the corona of GJ\,745\,B seems to be covered 
with X-ray dark CHs, a flare occurring in a single region capable to hold closed field lines has a great impact on the X-ray light curve and the coronal temperature (hardness ratio) of the star.
In fact, above we have shown that the time-averaged X-ray luminosity of GJ\,745\,B can be explained if a minor fraction of its  corona  ($0.72$\,\%) is covered with an eruptive solar-like CO.

By inference from our Sun, flares occur within COs. Assuming that the CO was already present before it erupted into a flare, a plausible description associates the {\it quiescent} corona of GJ\,745\,B  (before the flare) to a superposition of a quiescent CO and CHs. Equation~\ref{eq:ff} applied to $L_{\rm x,quies,GJ745B}$
yields an upper limit to the CO filling factor of $0.5$\,\%. 
To estimate the filling factor of the flare we take the maximum of the surface fluxes of solar CO from Table~\ref{tab:solar_regions} as a lower limit to the flux of the flare.
The peak X-ray luminosity during the flare ($\log{L_{\rm x,flare}}[\,erg/s] = 26.2$)  
then yields a 
filling factor of this flaring structure 
of at least $0.3$\,\%. 

\subsubsection{Metallicity and space motion}\label{subsubsect:discussion_CHstars_in_context_FeH}
\begin{figure}
    \centering
    \includegraphics[width=0.47\textwidth]{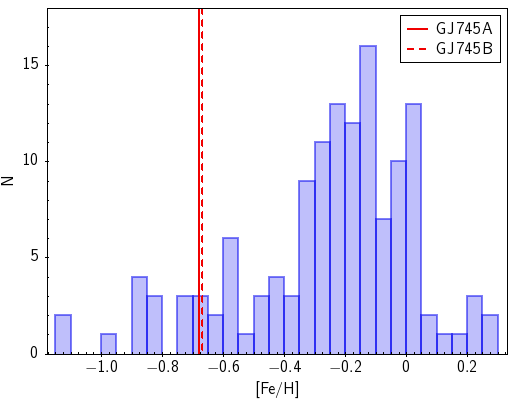}
    \caption{Metallicity of the {\sc M10pc-Gaia sample} extracted from the literature cited in Sect.~\ref{subsubsect:discussion_CHstars_in_context_FeH}. The coronal hole stars are highlighted.}
    \label{fig:hist_FeH}
    \end{figure}

In search for a parameter that distinguishes 
the CH M\,dwarfs
from the bulk of the {\sc M10pc-Gaia sample} we examined the metallicity from our compilation described in Sect.~\ref{sect:sample}. 
As can be seen from 
Fig.~\ref{fig:hist_FeH}, 
GJ\,745\,A and~B are, indeed, also
among the lowest metallicity stars 
in the {\sc M10pc-Gaia sample}.
Yet there are some other sample members with even lower values of [Fe/H] reported in the literature but at the high end of the X-ray brightness distribution. These are EV\,Lac, G\,19-7 and GJ\,867.   

Metallicity is difficult to measure in low-mass stars, and therefore it has long been a missing piece in the evolutionary picture. Recently, \cite{Amard20.0}  have presented evolutionary calculations which  predict that low-metallicity stars rotate faster at given age, yet they have a higher Rossby number because their convective turnover times are smaller. We caution that their model is valid for solar-like stars. If this scenario also holds for M\,dwarfs and if magnetic activity is ruled by Rossby number rather than rotation period alone,  low-metallicity stars are expected to be less active than solar- or higher-metallicity stars. \citet[][in their Fig.10]{Newton16.0} have examined the relation between [Fe/H] and $P_{\rm rot}$ and found  "no clear trend" for a sample of nearly $450$ M\,dwarfs. The lowest metallicities in their sample were, however, found at very slow rotators with rotation periods $P_{\rm rot} \approx 100$\,d. If the key parameter is $R_0$ this does not contradict the expectation of lowered activity in low-metallicity stars. In the study of \cite{Newton16.0} only two stars have such a low metallicity as
our CH stars.

We also computed the space motions of GJ\,745\,A and~B using their \textit{Gaia} coordinates, proper motions, parallaxes, and radial velocities. The result is given in Table~\ref{tab:CHstars}. 
GJ\,745\,A and~B are thin disk objects according to the Toomre diagram in \citet{2021ApJ...915L...6K}.

\subsubsection{Rotation}\label{subsubsect:discussion_CHstars_in_context_rotation}

Available studies on the stellar rotation indicate that GJ\,745\,A and~B are very slow rotators. Spectroscopic observations by e.g. \citet{1998A&A...331..581D, 2010AJ....139..504B} and \citet{2012AJ....143...93R} could only report upper limits on the rotational velocity $v\,sin\,i$. The most recent measurement by \citet{2018A&A...614A..76J} with CARMENES resulted in $v\,\sin{i}\,<\,2$\,km/s. Photometric studies using MEarth data \citep{2015ApJ...812....3W,Newton16.0} and a combination of MEarth and ASAS data \citep{2019A&A...621A.126D} yielded no detections for the rotation period. In fact, the latter work  reported a period of $P_{\mathrm{rot}}=3.8\,\pm\,0.01$\,d for GJ\,745\,B but with a high false alarm probability of $0.91$\,\%.

GJ\,745\,A and~B were observed  with the Transiting Exoplanet Survey Satellite (TESS) in sectors 40 and 54. We downloaded the two-minute  cadence light curves 
from the  MAST (Barbara A. Mikulski Archive for Space Telescopes) Portal. The search for the rotation period was carried out as explained by \citet{2022A&A...665A..30S} using three different methods: the generalized Lomb-Scargle-Periodogram \citep[GLS,][]{2009A&A...496..577Z}, autocorrelation function and sinefit. For each sector separately we searched for periods up to the duration of a TESS sector, which is about $27$\,days. All three methods for both stars in both sectors found signals with amplitudes that are between $3.5$ and $18$ times lower than the standard deviation of the light curves. Hence, we did not detect any rotation period for the two stars.

\subsubsection{Chromospheric activity}\label{subsubsect:discussion_CHstars_in_context_chromo}

Reports on chromospheric activity of GJ\,745\,A and~B include H$\alpha$ equivalent widths (EWs) of $0.06\,\AA$ for GJ\,745\,A \citep{Newton17.0} and around $-0.1\,\AA$ for GJ\,745\,A and~B \citep{2019A&A...623A..44S}. Hence, only a weak H$\alpha$ line is present meaning that GJ\,745\,A and~B are inactive stars.

To compare the H$\alpha$ EWs of GJ\,745\,A and~B to all stars in the {\sc M10pc-Gaia Sample} we collected H$\alpha$ EWs measurements from the literature. \citet{2014MNRAS.443.2561G} reported H$\alpha$ EWs for 129 M\,dwarfs from our sample. If available we updated this table with more recent measurements from \citet{Newton17.0,2019A&A...623A..44S,2020ApJ...905..107M} and \citet{2021ApJS..253...19Z}. Finally, we added values from \citet{2002AJ....123.3356G} and \citet{2012MNRAS.421.3189H} for four additional stars. In total, we found H$\alpha$ EWs for $136$ of the $150$ stars. The comparison with our full sample showed that GJ\,745\,A and~B are typical for the low-activity stars in the {\sc M10pc-Gaia sample}. The comparison between the coronal and the chromospheric emission of the {\sc M10pc-Gaia} stars will be discussed 
in more detail in a future work.


\section{Summary and conclusions}

We have explored the X-ray emission levels of M\,dwarfs on the volume-limited sample within $10$\,pc. Comparison with the flux emitted by individual magnetic structures on the corona of the Sun shows that the full range covered by these solar regions,  from emission brighter than the cores of active regions to the faint background Sun, is present in M\,dwarfs. The X-ray faintest M\,dwarfs comply with the faintest X-ray radiation of the Sun, suggesting that magnetic structures on M dwarf coronae are of the same nature as the ones observed on the Sun. Most notably, we have identified two stars that are fainter than even an hypothetic star entirely covered with average solar BKC.
The primary component of the common proper motion binary, GJ\,745\,AB, is undetected at an X-ray  flux level within the range displayed by solar CHs. Its companion is the faintest X-ray detected star in the whole {\sc M10pc-Gaia sample} and its X-ray surface flux is at the upper end of 
that of solar CHs. The flux of both components is compatible with a corona entirely covered by CHs or BKC, or a combinations of them: considering a dominating emission of CHs, the resulting filling factor of BKC is 21\% for GJ\,745\,A, while for GJ\,745\,B is 48\%. Another possible scenario is a corona covered by CHs, with a
small fraction of
ARs or even COs. In fact, our time-resolved analysis of 
GJ\,745\,B
has demonstrated that the detection is owed to a flare, while 
the star
is X-ray dark outside this event, just as its twin GJ\,745\,A. 
Specifically, we have computed a coronal filling factor with (flaring) CO of $0.3$\,\% for  GJ\,745\,B as being consistent with its X-ray flux during the flare. 

From this estimate for the CO filling factor we can make inferences on the geometry of the flare. 
Assuming the $0.3$\,\% CO filling factor  
to be concentrated in a single CO, this structure has an area of $2 \cdot 10^{19}\,{\rm cm^2}$. However, it is likely that the flaring area was a fraction of the CO area but with a higher flux. As a template we make use of the X9 flare on the Sun analyzed by \cite{Reale_01} which covered an area of $3 \cdot 10^{19}\,{\rm cm^2}$ and its flux was $\log{F_{\rm X9, SURF}}\,{\rm [erg/cm^2/s]} = 7.93$. A flare of the same flux combined with the rest of the corona of GJ\,745\,B being covered with CH is consistent with the observed X-ray flux of the star if the flare filling factor was $0.02$\,\%. We note that the solar X9 flare covered a similar area fraction, namely  $0.05$\,\% of the solar surface. However, due to the smaller radius of GJ\,745\,B, $0.02$\,\% surface coverage  corresponds to an area of only $1.8 \cdot 10^{18}\,{\rm cm^2}$. 
If we assume a loop of length $L$ and width $w \approx 0.1 L$ \citep{White02.0,Aschwanden17.0} this area translates into $L \approx 4 \cdot 10^9$\,cm, which is comparable to the typical loop lengths observed on solar flares \citep[see e.g. Table~2 in][]{Reale_01} and on the prototypical M dwarf flare star AD Leo \citep{Stelzer22.0}. 

An empirical relation between X-ray temperature and surface flux for coronally active stars was presented by \cite{Johnstone15.0}. If we assume that this relation holds also for individual flare events, for the solar X9 flare from \cite{Reale_01} a coronal temperature of $\gtrsim 10$\,MK is expected. From the simulated EPIC/pn spectra we found that the (time-averaged) hardness ratios of GJ\,745\,B suggest a temperature of $7.5$\,MK, in reasonable agreement with the temperature - flux relation given that this relation is as yet poorly calibrated, especially for M dwarfs. The systematic study of X-ray spectra, and hence coronal temperature, for the {\sc M10pc-Gaia sample} in a future work will put stronger constraints on this coronal scaling law.

We conclude from the X-ray properties of the faintest stars in the {\sc M10pc-Gaia sample} that the scenario for the structure of  coronae at the inactive tail of the M dwarf population is in accord with
a surface covered in large part by solar-like CHs, presumably dominated by open field lines, but with the possibility of individual active regions or cores of active regions that can produce the strongest signatures of activity, namely flares.

The extremely low X-ray activity of this binary -- outside episodic events -- is consistent with its other properties, like ultra-low metallicity, non-detection of photometric star spot variability, and low chromospheric emission,  which all suggest an old age.
Notably, both  GJ\,745 A and~B are {\it Gaia} Spectrophotometric Standard Stars, \citep{Pancino21.0} and {\it Gaia} radial velocity standard  stars \citep{Soubiran18.0},
testifying the absence of strong (activity-induced) variability in the optical waveband. 
The only property at odds with this evolutionary scenario are the space motions of GJ\,745\,A and~B, which place the binary within the Milky Way thin disk population.

\cite{Saar12.0} present some Maunder Minimum (MM) stars with $F_{\rm X,SURF}$ levels comparable to those we see in GJ\,745\,A and  GJ\,745\,B, that is in the range of solar CHs. They define a MM star through low  and little variable Ca\,{\sc ii}\,H\&K activity index, $R^{\prime}_{\rm HK}$, considering that the minimum value of $R^{\prime}_{\rm HK}$ depends on the metallicity of the star. The X-ray temperatures estimated by \cite{Saar12.0} for 
their MM 
stars from their {\it Chandra} hardness ratios are $T_{\rm x} \leq 1$\,MK, and thus expected to correspond to the upper boundary of the flux range for a CH we defined based on the literature. The MM stars in their sample are all  solar-type stars (SpT G). To our knowledge, no  dedicated study is present in the literature for MM-like M\,dwarfs. 

The particular importance of our study lies in the use of a volume-limited sample. This allows us to estimate the frequency of extremely inactive stars. In the whole {\sc M10pc-Gaia sample} of $141$ stars with sensitive X-ray data we have detected two such objects, hence an occurrence rate of $\sim 1$\,\%. The nine stars that still have to be observed within our {\it XMM-Newton} survey, are unlikely to significantly change this estimate. This result is of importance for exoplanet studies, as inactive stars offer the highest promise for habitability. GJ\,745 is a target of different exoplanet search samples, e.g. with CARMENES \citep[e.g.][]{2021A&A...656A.162M} and the Keck telescope \citep{2017AJ....153..208B}, so far with no success.
The binary is also among the targets of breakthrough listen search \citep{2017PASP..129e4501I}. Our systematic survey of the X-ray brightness of nearby M\,dwarfs allows us to identify the stars with faint high-energy emission, typical of solar BKC and CH, that might be the most suited for hosting life. While here we have focused on the X-ray dark stars among the M\,dwarfs within $10$\,pc, the   X-ray properties of the full {\sc M10pc-Gaia sample} will be presented in detail in our future work.

\begin{acknowledgements}
The authors would like to thank the anonymous referee for the useful and detailed comments to improve the manuscript.
MC acknowledges financial support by the Bundesministerium für Wirtschaft und Energie through the Deutsches Zentrum für Luft- und Raumfahrt e.V. (DLR) under grant number FKZ 50 OR 2105.
EM is supported by Deutsche Forschungsgemeinschaft under grant STE 1068/8-1. 
KP acknowledges support from the German \textit{Leibniz-Gemeinschaft} under project number P67/2018.
      This research has made use of data obtained from the 4XMM {\it XMM-Newton} serendipitous source catalog compiled by the $10$ institutes of the {\it XMM-Newton} Survey Science Centre selected by ESA, and of archival data of the ROSAT space mission. This work is also based on data from eROSITA, the soft X-ray instrument aboard SRG, a joint Russian-German science mission supported by the Russian Space Agency (Roskosmos), in the interests of the Russian Academy of Sciences represented by its Space Research Institute (IKI), and the Deutsches Zentrum für Luft- und Raumfahrt (DLR). The SRG spacecraft was built by Lavochkin Association (NPOL) and its subcontractors, and is operated by NPOL with support from the Max Planck Institute for Extraterrestrial Physics (MPE). The development and construction of the eROSITA X-ray instrument was led by MPE, with contributions from the Dr. Karl Remeis Observatory Bamberg \& ECAP (FAU Erlangen-Nuernberg), the University of Hamburg Observatory, the Leibniz Institute for Astrophysics Potsdam (AIP), and the Institute for Astronomy and Astrophysics of the University of Tübingen, with the support of DLR and the Max Planck Society. The Argelander Institute for Astronomy of the University of Bonn and the Ludwig Maximilians Universität Munich also participated in the science preparation for eROSITA.
      This research has made use of data and/or software provided by the High Energy Astrophysics Science Archive Research Center (HEASARC), which is a service of the Astrophysics Science Division at NASA/GSFC. 
      
\end{acknowledgements}

   \bibliographystyle{aa} 
   \bibliography{biblio} 

\begin{appendix} 
\section{X-ray  detection of GJ\,643}
\label{app:gj643}

The M3.5V dwarf GJ\,643 is located at only 74$^{\prime\prime}$ from GJ\,644\,AB, another member of the M10pc-Gaia sample with the same SpT, with which it forms a proper motion pair. GJ\,644\,AB is a spectroscopic binary and the B component is a spectroscopic binary as well \citep{Pourbaix_04}. An additional component of this multiple system is GJ\,644\,C, better known as vB\,8, which is spatially resolved from GJ\,644\,AB and GJ\,643, but with its SpT of M7 is not part of our sample.

\begin{figure}
\centering 
\includegraphics[width=0.40\textwidth]{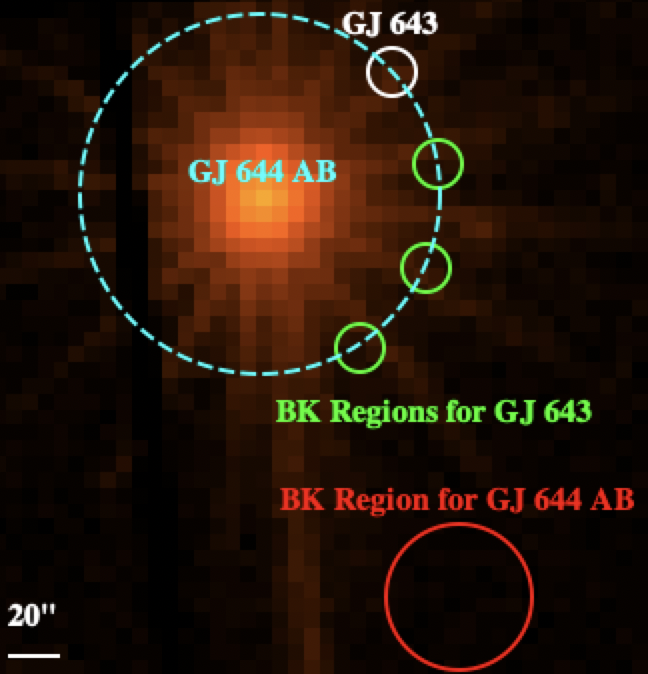}
\caption{EPIC/pn image of GJ\,643 and GJ\,644\,AB  (ObsID 0860302501). The dashed turquoise circle is centered on the X-ray position of  GJ\,644\,AB and its radius corresponds to the separation of GJ\,643 from that bright X-ray source. The three green circles are the background regions chosen for GJ\,643, the red circle represents  the background region for GJ\,644\,AB.}
\label{gj643_ds9}
\end{figure}

GJ\,644\,AB is a very bright X-ray source  (count rate of 12.7\,cts/s and $\log{L_{\rm x}}\,{\rm [erg/s]} =28.9$) with the wings of its Point Spread Function (PSF) extending beyond the position of GJ\,643 (see Fig.\ref{gj643_ds9}). The detection of a possible faint signal from GJ\,643, therefore, required a customized analysis. 

We analysed the {\it XMM-Newton} EPIC/pn observation using the Science Analysis Software (\texttt{SAS}) version 19.1.0 developed for the satellite. By examining the high energy events ($\geq 10$\,keV) across the full EPIC/pn detector, we excluded 
the time intervals affected by solar particle background. In addition we filtered the data for pixel patterns ($0 \le$ pattern $\le 12$), quality flag (flag = 0) and events channels ($PI \ge 150$). 
We performed  source detection in three energy bands: $0.2 - 0.5$\,keV (S), $0.5 - 1.0$\,keV (M), and $1.0 - 2.0$\,keV (H), after having removed the out-of-time events  caused by the intense emission of GJ\,644\,A, that could affect the position of the source in the image. 
Only the brighter star GJ\,644\,AB is detected.

To study GJ\,643, as visualized in Fig.~\ref{gj643_ds9}, we extracted the photons from a circular region of  $10^{\prime\prime}$ radius around the optical position of GJ\,643 which we determined by propagating the {\it Gaia} position with its proper motion  
to the epoch of the {\it XMM-Newton} observation. The background was extracted from three regions of 10$^{\prime\prime}$ radius located at the same distance from GJ\,644\,AB but at different angles, in order to take into account the contamination from GJ\,644\,AB. We also extracted the photons of GJ\,644\,AB from a $20^{\prime\prime}$ circular region around the detected X-ray source, choosing as background an adjacent circular region with radius of $30^{\prime\prime}$. 

\begin{figure}
\includegraphics[width=0.50\textwidth, angle=0] {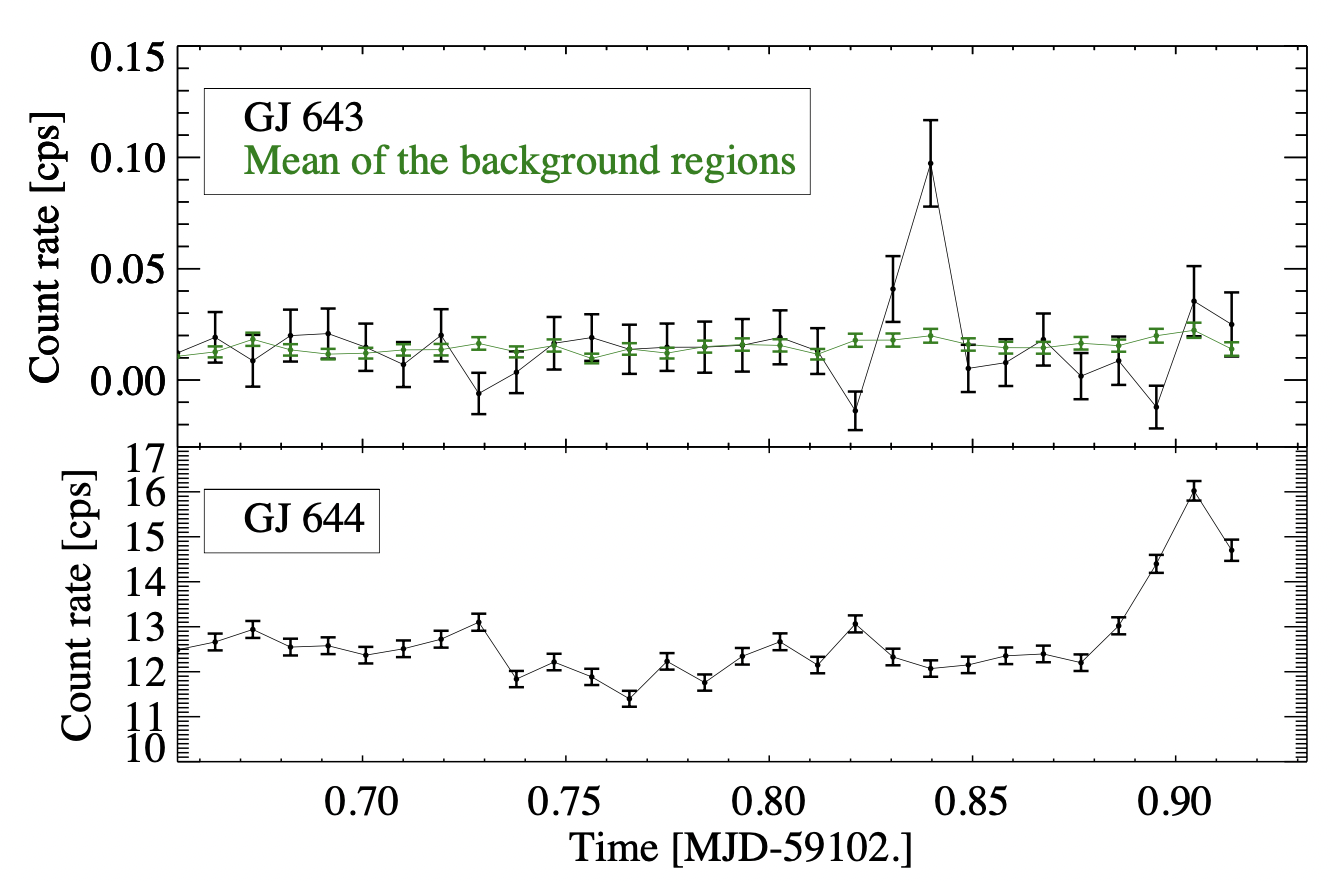}
\caption{Top panel shows EPIC/pn background subtracted lightcurve of GJ\,643 (black) and the lightcurve of its background (green), averaged over three regions (see Fig. \ref{gj643_ds9}) . Lower panel shows, for comparison, the lightcurve of GJ\,644\,AB.
}
\label{fig:lightcurve_GJ643}
\end{figure}

We then performed a temporal analysis 
at the two positions,
correcting the photon arrival times with the \texttt{SAS} tool \texttt{ barycen} and subtracting the individual backgrounds  of GJ\,643 and GJ\,644 with the \texttt{SAS} task \texttt{ epiclccorr}, which also corrects for  instrumental effects. 
The top panel of Fig.~\ref{fig:lightcurve_GJ643} shows the lightcurve of GJ\,643 together with the mean of its three background regions. This mean background has already been removed from the light curve of GJ\,643, thus eliminating the contamination by GJ\,644\,AB. This analysis unveils a flare on GJ\,643 and very weak quiescent emission. The mean background-subtracted count rate of the GJ\,643, calculated from the light-curve is $0.0159$\,cts/s,  corresponding to an X-ray luminosity of $\log{L_{\rm x}}\,{\rm[erg/s]} = 26.0$ that has to be interpreted as the 
emission of the X-ray source, averaged over the whole observation, without the contamination of GJ\,644\,AB.

In the bottom  panel of Fig.~\ref{fig:lightcurve_GJ643}
we report for comparison the light curve of GJ\,644\,AB.  
Clearly, the flare of GJ\,643 has no correspondence in the light curve of GJ\,644\,AB, evidencing that with our specific background subtraction we have achieved a reliable detection of GJ\,643. 

\end{appendix} 

\end{document}